\documentclass[a4paper,11pt]{article}
\usepackage{pos}
\usepackage{graphics}
\usepackage{float}

\usepackage[giveninits=true,sorting=none]{biblatex}
\usepackage[vlined]{algorithm2e}

\title{Fast Entropy Coding for ALICE Run 3}

\manuallySeparateAuthors
\author*[a]{Michael Lettrich} 
\author{for the ALICE collaboration}

\affiliation[a]{CERN, Technische Universit\"at M\"unchen,\\
                Geneva, Switzerland}

\emailAdd{michael.lettrich@cern.ch}

\abstract{
In LHC Run 3, the upgraded ALICE detector will record Pb-Pb collisions at a rate of 50 kHz using continuous readout. The resulting stream of raw data at ~3.5 TB/s has to be processed with a set of lossy and lossless compression and data reduction techniques to a storage data rate of ~90 GB/s  while preserving relevant data for physics analysis. This contribution presents a custom lossless data compression scheme based on entropy coding as the final component in the data reduction chain which has to compress the data rate from ~300 GB/s to ~90 GB/s. A flexible, multi-process architecture for the data compression scheme is proposed that seamlessly interfaces with the data reduction algorithms of earlier stages and allows to use parallel processing in order to keep the required firm real-time guarantees of the system. The data processed inside the compression process have a structure that allows the use of an rANS entropy coder with more resource efficient static distribution tables. Extensions to the rANS entropy coder are introduced to efficiently work with these static distribution tables and large but sparse source alphabets consisting of up to 25 Bit per symbol. Preliminary performance results show compliance with the firm real-time requirements while offering close-to-optimal data compression.

}
\FullConference{%
  40th International Conference on High Energy physics - ICHEP2020\\
  July 28 - August 6, 2020\\
  Prague, Czech Republic (virtual meeting)
}

\begin{document}
\maketitle

\section{Introduction}
\label{sec:introduction}

ALICE (A Large Ion Collider Experiment) \cite{alice} is a heavy-ion collision detector at the LHC (Large Hadron Collider) \cite{lhc} at CERN, built to study the physics of strongly interacting matter.
Throughout the Long Shutdown 2 (LS2) of the LHC, the ALICE detector receives a substantial upgrade \cite{alice_upgrade_letter} and will record Pb-Pb collisions at a rate of 50 kHz using continuous readout with an improved tracking precision in the upcoming Run 3 and Run 4 of the LHC. The resulting raw data rate of $\sim$3.5 TB/s needs to be decreased to a storage data rate of $\sim$90 GB/s. This is achieved by the ALICE Online-Offline (O$^{2}$) software \cite{alice_tdr} via a sequence of compression and data reduction steps without affecting physics. The final stage in this chain is a data compression scheme that provides a lossless, space efficient representation of the input data suitable for permanent storage.

General purpose compression schemes such as gzip/deflate \cite{deflate} and Zstandard \cite{zstd} are designed to provide good compression without prior knowledge of the processed data. Compression schemes that take into account the structure of the data however can be significantly more efficient as is shown e.g. by the Draco 3D data compression scheme \cite{draco} for 3D geometries or purpose built compression schemes for data acquisition systems (DAQ) \cite{duda2015designing}. 
Therefore ALICE in LHC Run 2 used a custom compression scheme based on the Huffman entropy coder \cite{tpcRun2} as well. However with a new approach to data taking and processing during LHC Run 3 as well as considering technological advances in compression algorithms, a completely new compression scheme has to be developed for Run 3.

The purpose of this contribution is thus to outline the main components of a custom data compression scheme for the ALICE detector in LHC Run 3. It describes the strategy used to compress the data from previous stages using rANS, a state of the art entropy coder and the required adaptations to rANS to allow fast and close-to-optimal entropy compression of ALICE Run 3 data.

\section{Choice of Compression Algorithm}
\label{sec:choice_of_compression_algorithm}
Data taking at 50 kHz continuous readout results in a stream of 3.5 TB/s, evenly split into time frames (TF) of $\sim$10--20 ms and distributed to $\mathcal{O}(250)$ Event Processing Nodes (EPN) such that each processes one TF at a time at firm real-time requirements. The result of zero suppression and lossy data reduction is a flat structure of integer arrays (SoA) which has to be compressed from $\sim$300 GB/s to $\sim$90 GB/s on the same EPN before being written to permanent storage as a compressed time frame (CTF) \cite{alice_tdr}. Each array inside an SoA has a defined value range of 4--25 Bits per value with additional padding and its own distribution of values. The length of the individual arrays however is variable and depends on the amount of extracted information from a raw time frame.

There are two major classes of widely used general purpose compression algorithms: dictionary compression and entropy compression. Both interpret the source data of a message $m$ as a concatenation of symbols $s_i$ from a finite alphabet $\mathcal{A}$, but rely on different concepts.
Dictionary compression replaces reoccurring sequences of symbols by a reference to a dictionary that is constructed by the algorithm on the fly. This principle is e.g. implemented in the LZ77, lzma and lz4 algorithms \cite{entropy_handbook}. Entropy coders on the other hand compress data based on the distribution of symbols in a message via a coding function $C$ that transforms source symbols into a representation where less probable symbols use more bits then highly probable symbols \cite{entropy_handbook}. Examples for entropy coders are Huffman coding \cite{huffman} and Asymmetric Numeral Systems coding (ANS) \cite{duda2009asymmetric}, \cite{duda2013asymmetric}.

General purpose compression schemes such as deflate (gzip) \cite{deflate} or the newer Zstandard \cite{zstd} combine both concepts by applying entropy compression on dictionary compressed data. The compression achieved by these schemes on simulated Run 3 data however was not satisfactory. It is highly likely that the probability for reoccurring patterns is small for our large alphabets of up to $2^{25}$ unique symbols and thus the dictionary compression is not effective. The entropy compression step in these schemes on the other hand cannot be adjusted sufficiently to our input data. For entropy coders compression performance does not depend on the size of the source alphabet $\mathcal{A}$ or reoccurring patterns but rather on a non-uniform distribution of source symbols. Therefore a plain entropy coder is the best choice for compression of ALICE Run 3 data.

The most suitable entropy coding algorithm for ALICE Run 3 data was selected in a study \cite{alirans} on simulated detector data of the ALICE time projection chamber (TPC). Evaluating compression rate and bandwidth as well as the ability to work with a $2^{25}$ Bit symbol alphabets, the rANS entropy coder, a variant of ANS, has shown the best and most consistent results across the input data. Given pre-calculated distribution tables for all arrays, a prototype rANS implementation in C++ managed to compress the contents of a SoA practically down to the bound of information-theory entropy $H$ \cite{shanon} achieving a compression factor 2 at an average bandwidth of 600 MB/s on commodity hardware. rANS was therefore selected for further investigation. With the lack of a universal library implementation of the algorithm however, an ALICE specific implementation is required.

\section{Entropy Coding Strategy}
\begin{figure}
  \centering
  \includegraphics[width=0.75\linewidth]{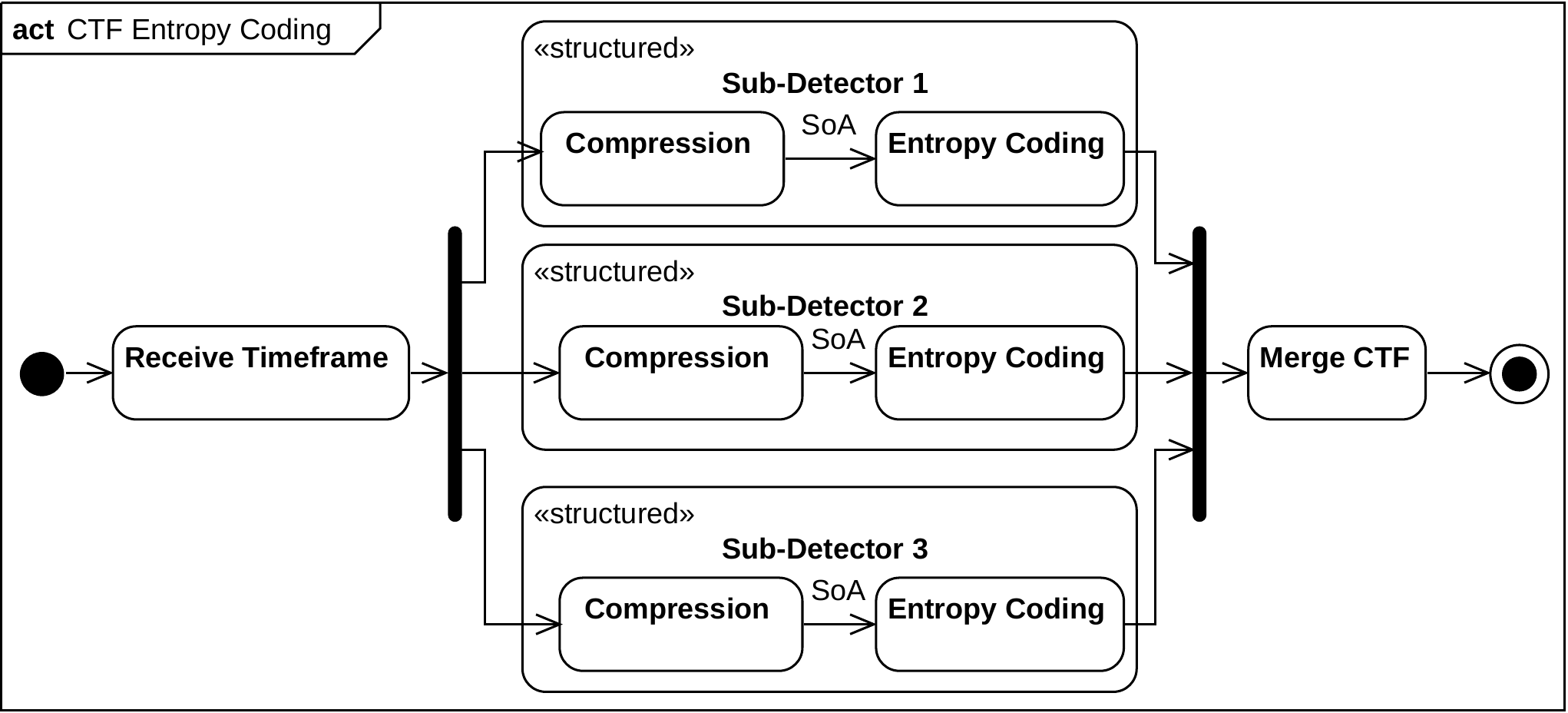}
  \caption{UML Activity diagram of the parallel, distributed processing of TF to CTF. Data from the TF is processed in a multi-stage compression and data reduction chain for each sub-detector producing a SoA which is entropy coded individually and merging into the CTF.}
  \label{fig:activityDiagram}
\end{figure}

The raw time frame is handled on the EPN by the ALICE O$^2$ data processing layer (DPL) \cite{dpl} --- a distributed, multi-process framework that allows connecting components via message passing. The SoAs constituting the CTF are produced in parallel by sets of multi-stage processes that compress the raw-data of one or multiple sub-detectors. Depending on the algorithms and the amount of data, the latency for each SoA is different. To prevent buffering of large amounts of data in shared memory, a distributed approach is chosen  where each SoA passes through its specific entropy coder before all fragments are merged into a final CTF that is sent to permanent storage (see Figure \ref{fig:activityDiagram}). The distributed approach furthermore decouples SoA specific pre-processing and entropy coding tasks from the final merging of uniformly structured blocks of encoded data.

The compression achievable by an entropy coder highly depends on how closely the distribution table used by the coder matches the underlying distribution of the input data. Individual compression of each array in the SoA respecting its value range and symbol distribution will yield the best results. Building the exact symbol distribution table for each array in each time-frame dynamically however is unfeasible as it would require a full pass over the input data before encoding can take place in a second pass which is too expensive in our setting. Additionally the information about the symbol distributions needs to be stored as metadata for decoding. The resulting increase in file size for source alphabets spanning a 25 Bit value range is not acceptable. However since a time-frame contains data of a large numbers of collisions, the distribution of the raw signals will not change unless the data-taking conditions change which will only happen over a time span of many time-frames. This allows pre-calculation of a distribution table for each individual array in a SoA respecting the specific value range and symbol distribution of the array and reuse the distribution table across time frames without heavy penalties on compression rate which was verified using simulated detector data. In addition the tables can be saved centrally to the ALICE Condition and Calibration Data Base (CCDB) \cite{alice_tdr} and fetched for decompression. This avoids large storage overhead caused by including distribution tables with each CTF file.

\section{Efficient Custom rANS Entropy Coder Implementation}
\label{sec:efficient_custom_rans_entropy_coder_implementation}
rANS is part of a family of variable range entropy coders called Asymmetric Numeral Systems (ANS) \cite{duda2009asymmetric} \cite{duda2013asymmetric}. Given a message $m$ consisting of symbols $s_i$ from a finite alphabet $\mathcal{A}$ and a probability distribution $f$, an arithmetic coding function $C_f:\mathcal{A} \mapsto \mathbb{N}$  encodes all symbols $s_i \in m$ into a single integer $x \in  \mathbb{N}$ called the state variable. Starting from an empty initial state $x_0$, symbol $s_i$ is encoded onto a state $x_{i-1}$ containing encoded information of all symbols $s_1,\dots,s_{i-1}$. This will lead to a new state $x_{i} = C_f(x_{i-1},s_i) > x_{i-1}$ that grows inversely proportional to the probability of the encoded symbol, i.e. $x_{i}\approx x_{i-1}/Pr[s_i]$. Renormalization keeps $x$ constrained within an interval $I$ that can efficiently be handled by a computer and bits are streamed out if the upper limit is surpassed during encoding or read in when the lower limit is surpassed during decoding. The state variable $x$ behaves like a last-in-first-out (LIFO) stack which requires the decoder to always exactly invert the encoding step $D(C(x_i,s_i))=(x_i,s_i)$ to recover the input. The generalization of this idea is that, an arbitrary transformation $t$ can be applied on to state $x$ during encoding as long as it is inverted by $t^{-1}$ during decoding, which also allows nesting i.e. $t_n^{-1}(\dotsc t_1^{-1}(D_f(C_f(t_1(\dotsc t_n(x_i,s_i))))))=(x_i,s_i)$. Efficient implementations on pipelined, SIMD capable CPUs or GPGPUs \cite{giesen2014interleaved} rely on these transformations to enable instruction level parallelism.

The ALICE rANS implementation additionally uses a transformation function $t$ for handling static distribution tables. With larger alphabets chances increase to encounter infrequent symbols with a probability close to zero. The pre-calculated distribution table thus can contain $Pr[s_r]=0$ for a rare symbol $s_r$ which is incompatible with the rANS algorithm, that strictly requires $Pr[s_i]>0, \forall s_i \in \mathcal{A}$. Incompressible symbols can be encoded by introducing a functional symbol $r$ into $\mathcal{A}$ with $Pr[r]>0$. If a symbol $s_i$ is marked as incompressible in the distribution table, a transformation replaces $s_i$ with $r$ and passes it to the encoder. The original symbol $s_i$ is pushed onto a stack which is appended as a special block at the end of the encoded data. If during decoding the functional symbol $r$ is encountered, it is replaced with the top element of the stack saved alongside the data. Algorithm \ref{algo:encode} and Algorithm \ref{algo:decode} formally describe the encoding/decoding of incompressible symbols. Run-length encoding (RLE) \cite{entropy_handbook} is implemented as a transformation in a similar way.

rANS relies on some costly arithmetic operations that depend on the probability of the encoded symbol. A pre-calculated lookup table (LUT) can be used to replace these reoccurring arithmetics with table lookups. For large alphabets with up to $2^{25}$ symbols these LUTs no longer fit into CPU cache, reducing the performance benefits. Thankfully, many of the distribution tables for these large alphabets are sparse, containing over 90\% incompressible symbols. Using a LUT with a single indirection instead of direct indexed lookup allows the implementation of more efficient data structure. Referencing all incompressible symbols directly to the special functional symbol $r$ shrinks sparse LUTs by up to a factor of 16 preventing cache eviction. The probability of a symbol directly translates to the expected frequency of lookup. Sorting symbols in storage by their probability measurably increases the probability of cache hits in higher level caches. Since the LUTs are reused for many time-frames, setup costs occur only during initialization.

\begin{minipage}[t]{.47\textwidth}
\centering
\vspace{0pt}  
\begin{algorithm}[H]
\SetAlgoLined
  \eIf{Pr[$x_i$]$>0$}{
   $C(x_i, s_i)$\;
   }{
   incompressible.push($x_i$)\;
   $C(x_i, r)$\;
  }
  \label{algo:encode}
 \caption{Encoder with\\ incompressible symbols}
\end{algorithm}
\end{minipage}%
\begin{minipage}[t]{.47\textwidth}
\centering
\vspace{0pt}  
\begin{algorithm}[H]
\SetAlgoLined
    $ s_i \leftarrow D(x_i)$ \;
  \eIf{$s_i == r$}{
     \Return incompressible.pop()\;
   }{
     \Return $s_i$\;
   }
   \label{algo:decode}
 \caption{Decoder with\\ incompressible symbols}
\end{algorithm}
\end{minipage}

\section{Status of the Implementation and Outlook}
The entropy compression scheme for ALICE Run 3 consists of two components, a general purpose, configurable rANS entropy coding library and an ALICE specific component performing compression of the SoAs and final CTF creation inside ALICE O$^2$ using the rANS library. At the time of writing a base implementation for both components exists and most of the sub-detectors are integrated. Preliminary measurements based on simulated detector data show excellent compression of SoAs by the entropy coder, within per mills to the information-theory limit of entropy $H$ \cite{shanon} while keeping the firm real-time requirements. For the production code further performance improvements can be achieved with a better use of pipeling, SIMD vectorization and multithreading. Optimizations in the ROOT based CTF data format can additionally decrease overhead introduced by metadata.

\section{Conclusion} 

The new, purpose build compression scheme presented in this contribution allows the ALICE O$^2$ framework to reduce the amount of data sent to storage effectively. Combining the flexibility of the O$^2$ DPL with a custom implementation of a rANS entropy coder that leverages the structure of the data allows fast and quasi-optimal compression while operating within the firm real-time bounds required by the online processing for ALICE in LHC Run 3. 

\printbibliography

\end{document}